\documentclass[12pt]{article}
\usepackage{graphicx}
\usepackage{graphics}
\title{Hypergeometric summation representations of the Stieltjes constants} 
\author{Mark W. Coffey\\
Department of Physics\\
Colorado School of Mines\\
Golden, CO  80401\\
(Received $\mbox{~~~~~~~~~~~~~~~~~~~~~~~~~~~~~~~2011}$)}
\date{March 9, 2011}
\begin{document}
\maketitle
\baselineskip=25 pt
\begin{abstract}

The Stieltjes constants $\gamma_k$ appear in the regular part of the Laurent
expansion of the Riemman and Hurwitz zeta functions.  We demonstrate that these
coefficients may be written as certain summations over mathematical constants
and specialized hypergeometric functions $_pF_{p+1}$.  This family of results
generalizes a representation of the Euler constant in terms of a summation over
values of the trigonometric integrals Si or Ci.  The series representations are
suitable for acceleration.  As byproducts, we evaluate certain sine-logarithm
integrals and present the leading asymptotic form of the particular $_pF_{p+1}$
functions. 

\end{abstract}
 
\medskip
\baselineskip=15pt
\centerline{\bf Key words and phrases}
\medskip 

\noindent

Riemann zeta function, Stieltjes constants, generalized hypergeometric function, Gamma function, digamma function, Euler constant, series representation, integral representation, Hurwitz zeta function, cosine integral, sine integral 

\vfill
\centerline{\bf 2010 AMS codes} 
11M06, 11M35, 11Y60, 33C20

\baselineskip=25pt
\pagebreak
\medskip
\centerline{\bf Introduction and statement of results}
\medskip

Recently we developed series representations of the Euler constant $\gamma$ and
values of the Riemann zeta function at integer argument, together with other mathematical
constants, in terms of summations over the trigonometric integrals Si and Ci
\cite{coffeydigammarepr} (Propositions 2 and 3, Corollaries 7 and 9).  The present work is in a sense a generalization of those results.  We present series representations of the Stieltjes (generalized Euler) constants that involve sums over certain generalized
hypergeometric functions $_pF_{p+1}$.  There are underlying connections to Si and Ci
and certain logarithmic integrals of those functions, and our presentation includes
some developments of special function theory.

We let $\zeta(s)=
\zeta(s,1)$ be the Riemann zeta function \cite{edwards,ivic,riemann,titch}, $\Gamma$ the 
Gamma function, $\psi=\Gamma'/\Gamma$ be the digamma function (e.g., \cite{nbs}) with $\gamma=-\psi(1)$ the Euler constant, $\psi^{(k)}$ be the polygamma functions \cite{nbs}, and $_pF_q$ be the generalized hypergeometric function \cite{andrews}.  Although we
concentrate on the Stieltjes constants $\gamma_k=\gamma_k(1)$ corresponding to the
Riemann zeta function, we briefly describe how the approach carries over to those for
the Hurwitz zeta function.  (See the discussion section.) 
The Stieltjes constants $\gamma_k(a)$ \cite{coffeyjmaa,coffey2009,coffeyjnt,coffeystdiffs,matsuoka,matsuoka2,
stieltjes,wilton2} arise in the regular part of the Laurent expansion of the Hurwitz zeta function $\zeta(s,a)$:
$$\zeta(s,a)={1 \over {s-1}}+\sum_{n=0}^\infty {{(-1)^n} \over {n!}}\gamma_n(a)(s-1)^n,
\eqno(1.1)$$
where $\gamma_0(a)=-\psi(a)$.

The Stieltjes constants may be expressed through the limit relation \cite{berndt}
$$\gamma_n(a)={{(-1)^n} \over {n!}} \lim_{N\to \infty}\left[\sum_{k=0}^N {{\ln^n (k+a)}
\over {k+a}}-{{\ln^{n+1} (N+a)} \over {n+1}}\right], ~~~~~~n \geq 0.  \eqno(1.2)$$
Here, $a \notin \{0,-1,-2,\ldots\}$.  The sequence $\{\gamma_k(a)\}_{k \geq 0}$
has rapid growth in magnitude with $k$ for $k$ large and changes in sign due to both
$k$ and $a$.  Subsequences of the same sign of arbitrarily long length occur.
For an asymptotic expression for these constants,
even valid for moderate values of $k$, \cite{knessl2} (Section 2) may be consulted.
The Stieltjes constants appear in applications including asymptotic analysis whether
in computer science or high energy physics.

{\bf Proposition 1}.  (a)
$$\gamma_1=2\sum_{n=1}^\infty \left[1+{1 \over 2}(\gamma-2)\gamma-{\pi^2 \over {24}}
-{{\pi^2 n^2} \over 6} ~_3F_4\left(1,1,1;2,2,2,{5 \over 2};-\pi^2 n^2\right) \right.$$
$$\left. +\ln(2\pi n)\left(\gamma-1+{1 \over 2}\ln(2\pi n)\right)\right]+{1 \over 2}-\gamma, \eqno(1.3)$$
and (b) the slightly accelerated form
$$\gamma_1=2\sum_{n=1}^\infty \left[1+{1 \over 2}(\gamma-2)\gamma-{\pi^2 \over {24}}
-{5 \over {32\pi^4}}{1 \over n^4}
-{{\pi^2 n^2} \over 6} ~_3F_4\left(1,1,1;2,2,2,{5 \over 2};-\pi^2 n^2\right) \right.$$
$$\left. +\ln(2\pi n)\left(\gamma-1+{1 \over 2}\ln(2\pi n)\right)\right]+{{73} \over {144}}-\gamma,  \eqno(1.4)$$
and (c)
$$\gamma_2=1-2(\gamma+\gamma_1)+2\sum_{n=1}^\infty \left\{2(1-\gamma)+\gamma^2+{\gamma^3
\over 3}-{\pi^2 \over {12}} +\gamma {\pi^2 \over {12}}-{2 \over 3}\zeta(3) \right.$$
$$+{{\pi^2 n^2} \over 6} ~_4F_5\left(1,1,1,1;2,2,2,2,{5 \over 2};-\pi^2 n^2\right)$$
$$\left.
-2\ln(2 \pi n)\left[1-\gamma+{\gamma^2 \over 2}-{\pi^2 \over {24}}+{1 \over 2}\ln(
2\pi n)(\gamma-1)+{1 \over 6}\ln^2(2\pi n)\right]\right\}.  \eqno(1.5)$$

{\it Remark}.  The summand in (1.5) is $O(n^{-3})$ as $n \to \infty$.  All parts of the
Proposition may be further accelerated in their rate of convergence.

{\bf Proposition 2}.  The Stieltjes constant $\gamma_j$ may be expressed as
a summation over $n$ of mathematical constants, terms $\ln^k(2\pi n)$, with $k=1,\ldots,
j$, and the term
$$\pi^2 \sum_{n=1}^\infty n^2 ~_{j+2}F_{j+3}\left(1,1,\ldots,1;2,2,\ldots,2,{5 \over 2};
-\pi^2 n^2\right).  \eqno(1.6)$$

The following section of the paper contains the proof of the Propositions.  Section
3 contains various supporting and reference Lemmas.  Some of these Lemmas present 
results of special function theory and may indeed be of occasional interest in themselves.  Certain logarithmic integrals of the Si and Ci functions are considered in the Appendix.


\medskip
\centerline{\bf Proof of Propositions}


We let $P_1(x)=B_1(x-[x])=x-[x]-1/2$ be the first periodic Bernoulli polynomial, with $\{x\}=x-[x]$ the fractional part of $x$.  Being periodic, $P_1$ has 
the Fourier series \cite{nbs} (p. 805)
$$P_1(x)=-\sum_{j=1}^\infty {{\sin 2\pi j x} \over {\pi j}}.  \eqno(2.1)$$

{\it Proposition 1}.
(a) We take $a=1$ in the representation 
$$\zeta(s,a)={a^{-s} \over 2}+{a^{1-s} \over {s-1}}-s\int_0^\infty {{P_1(x)} \over
{(x+a)^{s+1}}} dx, ~~~~\sigma \equiv \mbox{Re} ~s >0, \eqno(2.2)$$
so that
$$\zeta(s)={1 \over 2}+{1 \over {s-1}}-s\int_1^\infty {{P_1(x)} \over {x^{s+1}}} dx, ~~~~\sigma >-1, \eqno(2.3)$$ 
and
$$\zeta'(s)+{1 \over {(s-1)^2}}=-\int_1^\infty {{P_1(x)} \over x^{s+1}}dx+s \int_1^
\infty {{P_1(x)} \over x^{s+1}} \ln x ~dx.  \eqno(2.4)$$
Taking $s\to 1$ in (2.4) and using (1.1) gives
$$\gamma_1=-\int_1^\infty {{P_1(x)} \over x^2} \ln x ~dx+{1 \over 2}-\gamma.  \eqno(2.5)$$
Here, we have used the well known integral that also results from (2.3)
$$\int_1^\infty {{P_1(x)} \over x^2}dx=\sum_{j=1}^\infty \int_j^{j+1} {{(x-j-1/2)} \over
x^2}dx$$
$$=\sum_{j=1}^\infty \left[\ln\left({{j+1} \over j}\right)-{1 \over 2}\left({1 \over {j+1}}+{1 \over j}\right)\right]={1 \over 2}-\gamma.  \eqno(2.6)$$
From Lemmas 2, 3, and 6 we have
$${1 \over \kappa}\int_1^\infty {{\sin \kappa x} \over x^2}\ln x ~dx=1+{1 \over 2}(\gamma
-2)\gamma-{\pi^2 \over {24}}-{\kappa^2 \over {24}} ~_3F_4\left(1,1,1;2,2,2,{5 \over 2};
-{\kappa^2 \over 4}\right)$$
$$+{{\ln \kappa} \over 2}(2\gamma-2 +\ln \kappa).  \eqno(2.7)$$
Taking $\kappa=2\pi n$ and using (2.1) and (2.5) so that
$$\gamma_1={1 \over \pi} \sum_{n=1}^\infty {1 \over n}\int_1^\infty {{\sin(2\pi n x)}
\over x^2}\ln x ~dx +{1 \over 2}-\gamma, \eqno(2.8)$$
gives the first part of the Proposition from (2.7).

(b) 
The summand of (1.3) is $O(n^{-4})$ as $n \to \infty$.  By adding and subtracting
${5 \over {32}}\sum_{n=1}^\infty {1 \over n^4}={5 \over {32}}{{\zeta(4)} \over \pi^4}
={1 \over 288}$ we transform the summand to be $O(n^{-6})$ as $n \to \infty$.

(c) From (2.4) we obtain
$$\zeta''(s)-{2 \over {(s-1)^3}}=2\int_1^\infty {{P_1(x)} \over x^{s+1}}\ln x ~dx-s \int_1^\infty {{P_1(x)} \over x^{s+1}} \ln^2 x ~dx.  \eqno(2.9)$$
Then from (1.1)
$$\gamma_2=\int_1^\infty {{P_1(x)} \over x^2}(2-\ln x)\ln x ~dx, \eqno(2.10)$$
where by (2.5) the term
$$2\int_1^\infty {{P_1(x)} \over x^2}\ln x ~dx=1-2(\gamma+\gamma_1).  \eqno(2.11)$$
For the term
$$\int_1^\infty {{P_1(x)} \over x^2}\ln^2 x ~dx=-{1 \over \pi} \sum_{n=1}^\infty {1 
\over n}\int_1^\infty {{\sin(2\pi n x)} \over x^2}\ln^2 x ~dx, \eqno(2.12)$$
we apply Lemmas 3, 4, and 6.  We omit further details.  

{\it Proposition 2}.  As may be proved from (2.2) by induction, for integers $j \geq 1$
we have
$$\zeta^{(j)}(s,a)=(-1)^j a^{1-s}\sum_{k=0}^j {j \choose k}(j-k)! {{\ln^k a} \over {(s-1)^{j-k+1}}}+{{(-1)^j} \over 2}a^{-s} \ln^j a$$
$$+(-1)^j \int_0^\infty {{P_1(x)} \over {(x+a)^{s+1}}}\ln^{j-1} (x+a)\left[j-s\ln(x+a)\right] ~dx.  \eqno(2.13)$$
At $a=1$ we have
$$\zeta^{(j)}(s)={{(-1)^j j!} \over {(s-1)^{j+1}}}
+(-1)^j \int_1^\infty {{P_1(x)} \over {x^{s+1}}}\ln^{j-1}x\left(j-s\ln x\right) ~dx.  \eqno(2.14)$$
Then by (1.1) we have
$$\gamma_j=\int_1^\infty {{P_1(x)} \over x^2} \ln^{j-1}x(j-\ln x)dx$$
$$=-{1 \over \pi} \sum_{n=1}^\infty {1 \over n}\int_1^\infty {{\sin(2\pi nx)} \over x^2}
\ln^{j-1}x(j-\ln x)dx.  \eqno(2.15)$$
Herein on the right side, $j$ multiplies a contribution from $\gamma_{j-1}$.
We then appeal to Lemmas 2, 3, and 4 for logarithmic-sine integrals, and
the result follows.

\medskip
\centerline{\bf Lemmas}

Herein the cosine integral is defined by
$$\mbox{Ci}(z)\equiv-\int_z^\infty {{\cos t} \over t}dt=\gamma+\ln z+\int_0^z {{\cos t-1}
\over t}dt.  \eqno(3.1)$$
As usual, $(w)_n=\Gamma(w+n)/\Gamma(w)$ denotes the Pochhammer symbol.  

{\bf Lemma 1}.  (Hypergeometric form of the cosine integral)
$$\mbox{Ci}(z)=\gamma+\ln z-{z^2 \over 4} ~_2F_3\left(1,1;2,2,{3 \over 2};-{z^2 \over 4}
\right).  \eqno(3.2)$$

{\it Proof}.  This easily follows from the expression
$$\mbox{Ci}(z)=\gamma+\ln z +\sum_{\ell=1}^\infty {{(-1)^\ell z^{2\ell}} \over {2\ell
(2\ell)!}}.  \eqno(3.3)$$

{\bf Lemma 2}.  (a) For $a \neq 0$,  
$$\int_x^y {{\mbox{Ci}(az)} \over z}dz=\gamma \ln(y/x)+{1 \over 2}[\ln^2(ay)-\ln^2(ax)]$$
$$-{a^2 \over 8}\left[y^2 ~_3F_4\left(1,1,1;2,2,2,{3 \over 2};-{{a^2 y^2} \over 4}\right) -x^2 ~_3F_4\left(1,1,1;2,2,2,{3 \over 2};-{{a^2 x^2} \over 4}\right)\right],  \eqno(3.4)$$
and (b) for $b>a>0$
$$\int_a^b {{\sin \kappa x} \over x^2}\ln x ~dx=\kappa\left\{\mbox{Ci}(\kappa
b)(1+\ln b)-\mbox{Ci}(\kappa a)(1+\ln a)-\gamma \ln(b/a) \right.$$
$$\left.+{1 \over 2}[\ln^2(\kappa a)-\ln^2(\kappa b)] + {\kappa^2 \over 8}\left[b^2 ~_3F_4\left(1,1,1;2,2,2,{3 \over 2};-{{\kappa^2 b^2} \over 4}\right) -a^2 ~_3F_4\left(1,1,1;2,2,2,{3 \over 2};-{{\kappa^2 a^2} \over 4}\right)\right]
\right\}$$
$$+{{\sin \kappa a} \over a}(1+\ln a)-{{\sin \kappa b} \over b}(1+\ln b).
\eqno(3.5)$$

{\it Proof}.  (a) We first note that $1/(\ell+1)=(1)_\ell/(2)_\ell$ and $(2\ell+2)!=\Gamma(2\ell+3)=2 \cdot 4^\ell(3/2)_\ell (\ell+1)\ell!$.  
The latter relation follows from the duplication formula 
$\Gamma(2j+1)=4^j \Gamma(j+1/2)j!/\Gamma(1/2)$.  
Then by using (3.3) we have
$$\int_x^y {{\mbox{Ci}(az)} \over z}dz=\int_x^y [\gamma +\ln(az)]{{dz} \over z}
+\sum_{\ell=1}^\infty {{(-1)^\ell a^{2\ell}}\over {2\ell (2\ell)!}}\int_x^y z^{2\ell-1}dz$$
$$=\gamma \ln(y/x)+{1 \over 2}[\ln^2(ay)-\ln^2(ax)]+\sum_{\ell=1}^\infty {{(-1)^\ell a^{2\ell}}\over {(2\ell)^2 (2\ell)!}}(y^{2\ell}-x^{2\ell}).  \eqno(3.6)$$
The sums are then rewritten according to
$$\sum_{\ell=1}^\infty {{(-1)^\ell a^{2\ell}}\over {(2\ell)^2 (2\ell)!}}y^{2\ell}
=-{1 \over 4}\sum_{\ell=0}^\infty {{(-1)^\ell a^{2\ell+2}}\over {(\ell+1)^2 (2\ell+2)!}} y^{2\ell+2}$$
$$=-{a^2 \over 8}\sum_{\ell=0}^\infty (-1)^\ell {{(1)_\ell^2} \over {(2)_\ell^2}}
{a^{2\ell} \over 4^\ell}{1 \over {(3/2)_\ell}}{{(1)_\ell} \over {(2)_\ell}}{y^{2\ell+2}
\over {\ell !}}$$
$$=-{a^2 \over 8}y^2 ~_3F_4\left(1,1,1;2,2,2,{3 \over 2};-{{a^2 y^2} \over 4}\right).
\eqno(3.7)$$

(b) We first have, by integrating by parts,
$$\int_a^b {{\sin \kappa x} \over x^2}\ln x ~dx=\int_a^b \left[{{\sin \kappa
x} \over x}+\kappa \cos \kappa x \ln x\right]{{dx} \over x}-\left.{{\sin \kappa x}\over x}\ln x\right|_a^b$$
$$=\kappa \int_a^b {{\cos \kappa x} \over x}\ln x ~dx+\kappa [\mbox{Ci}(
\kappa b)-\mbox{Ci}(\kappa a)]+{{\sin \kappa a} \over a}(1+\ln a)-{{\sin \kappa b} \over b}(1+\ln b).  \eqno(3.8)$$
Next, also by integration by parts,
$$\int_a^b {{\cos \kappa x} \over x}\ln x ~dx=-\int_a^b {{\mbox{Ci}(\kappa x)}\over x}dx +\mbox{Ci}(\kappa b)\ln b-\mbox{Ci}(\kappa a)\ln a.  \eqno(3.9)$$
We then apply part (a) to (3.9) and combine with (3.8).
   
{\bf Lemma 3}.  (a) For $j \geq 1$ an integer,
$$\int_1^\infty {{\sin \kappa x} \over x^2}\ln^j x ~dx=j\int_1^\infty {{\ln^{j-1} t}
\over t}\left[-\kappa \mbox{Ci}(\kappa t)+{{\sin \kappa t} \over t}\right]dt.  \eqno(3.10)$$
(b) With $j=1$,
$$\int_1^\infty {{\sin \kappa x} \over x^2}\ln x ~dx =-\kappa\left[\int_1^\infty
{{\mbox{Ci}(\kappa t)} \over t}dt + \mbox{Ci}(\kappa)\right]+ \sin \kappa.  \eqno(3.11)$$

{\it Proof}.  We interchange a double integral on the basis of Tonelli's theorem,
$$\int_1^\infty {{\sin \kappa x} \over x^2}\ln^j x ~dx=j\int_1^\infty {{\sin \kappa x} \over x^2} \int_1^x {{\ln^{j-1} t} \over t}dt$$
$$=j\int_1^\infty {{\ln^{j-1} t} \over t}dt \int_t^\infty {{\sin \kappa x} \over x^2}
dx$$
$$=j\int_1^\infty {{\ln^{j-1} t}
\over t}\left[-\kappa \mbox{Ci}(\kappa t)+{{\sin \kappa t} \over t}\right]dt,  \eqno(3.12)$$
with the aid of an integration by parts.  (b) follows easily.

Part (d) of the next result gives a prescription for successively 
obtaining certain needed $\log$-sine integrals, while part (b) complements
Lemma 3.
{\newline \bf Lemma 4}.  (a)
$$\int_1^\infty {{\cos(b x)} \over x}\ln x ~dx=-\int_1^\infty {{\mbox{Ci}
(bx)} \over x}dx, \eqno(3.13)$$
(b)
$$\int_1^\infty {{\sin(bx)} \over x^2}\ln x ~dx=\int_1^\infty \left[-b
\mbox{Ci}(bx)+{{\sin(bx)} \over x}\right]dx, \eqno(3.14)$$
(c)
$$\int_1^\infty {{\cos(b x)} \over x}\ln^j x ~dx=-j\int_1^\infty {{\mbox{Ci}
(bx)} \over x}\ln^{j-1}x ~dx, \eqno(3.15)$$
(d) Let
$$f_j(b)=\int_1^\infty {{\mbox{Ci} (bx)} \over x}\ln^j x ~dx
=\int_b^\infty {{\mbox{Ci}(y)} \over y}\ln^j\left({y \over b}\right)dy,
\eqno(3.16)$$
with $f_j(\infty)=0$, such that
$$-j\int_0^b f_{j-1}(b)db=\int_1^\infty {{\sin(bx)} \over x^2}\ln^j x ~dx
\equiv g_j(b), \eqno(3.17)$$
with $g_j(0)=g_j(\infty)=0$ and $g_0(b)=\sin b - b \mbox{Ci}(b)$.  Then for $j \geq 1$
$$f_j(b)=-j \int{{f_{j-1}(b)} \over b}db,  \eqno(3.18)$$
and
$$g_j(b)=-jb\int{{g_{j-1}(b)} \over b^2}db+c_jb,  \eqno(3.19)$$
where $c_j$ is a constant.  
{\newline (e) Let for $j \geq 1$}
$$h_j(b)=\int_1^\infty f(bx)\ln^j x ~dx={1 \over b}\int_b^\infty f(y)
\ln^j\left({y \over b}\right)dy, \eqno(3.20)$$
where $f(\infty)=0$ and $f \to 0$ sufficiently fast at infinity for the
integral to converge.  Then
$$h_j(b)=-{j \over b}\int h_{j-1}(b) db.  \eqno(3.21)$$

{\it Proof}. For (a) and (c) we integrate by parts, using that Ci$(x)$ is O$(1/x)$ as $x \to \infty$:
$$\int_1^\infty {{\cos(b x)} \over x}\ln^j x ~dx=-j\int_1^\infty {{\mbox{Ci}
(bx)} \over x} \ln^{j-1} x ~dx+\left.\mbox{Ci}(bx) \ln^j x \right|_1^\infty$$
$$=-j\int_1^\infty {{\mbox{Ci}(bx)} \over x} \ln^{j-1} x ~dx.  \eqno(3.22)$$
For (b) we use (a) and
$$\int_0^b \mbox{Ci}(bx)db=b\mbox{Ci}(bx)-{{\sin(bx)} \over x}, \eqno(3.23)$$
so that
$$\int_1^\infty {{\sin(bx)} \over x^2}\ln x ~dx=\int_0^b \int_1^\infty
{{\cos(bx)} \over x}\ln x ~dxdb$$
$$=-\int_0^b \int_1^\infty {{\mbox{Ci}(bx)} \over x}dx db.  \eqno(3.24)$$  
For (d) we use the property
$${\partial \over {\partial b}}\ln^j \left({y \over b}\right)=-{j \over b}
\ln^{j-1}\left({y \over b}\right).  \eqno(3.25)$$
Then for $j \geq 1$
$$f_{j-1}(b)=\int_b^\infty {{\mbox{Ci}(y)} \over y}\ln^{j-1}\left({y \over b}\right)dy$$
$$=-{b \over j}\int_b^\infty {{\mbox{Ci}(y)} \over y}{\partial \over {\partial b}}\ln^j\left({y \over b}\right)dy$$
$$=-{b \over j}{\partial \over {\partial b}}\int_b^\infty {{\mbox{Ci}(y)} \over y}\ln^j\left({y \over b}\right)dy$$
$$=-{b \over j}{\partial \over {\partial b}}f_j(b), \eqno(3.26)$$
from which (3.15) follows.  For (3.16), we have
$$g_j(b)=b\int_b^\infty {{\sin y} \over y^2} \ln^j\left({y \over b}\right)dy,
\eqno(3.27)$$ 
from which follows
$${\partial \over {\partial b}}g_j(b)={1 \over b}g_j(b)-{j \over b}g_{j-1}(b).  \eqno(3.28)$$
The solution of this linear differential equation gives (3.19).  That
$g_j(\infty)=0$ follows from the Riemann-Lebesgue Lemma applied to (3.17).
For (e) we have $h_{j-1}(b)=-(b/j)\partial_b h_j(b)$ from which (3.21) follows.

{\bf Corollary 1}.  By iteration of (3.19), we see that $g_j(b)$ 
contains a term with ${{(-1)^{j+1}} \over {j+1}}b\ln^{j+1} b$.

{\it Remark}.  
Imposing $g_1(\infty)=0$ gives $c_1=\gamma^2/2-\pi^2/24$. 

We note that Lemma 5 just below and the like allows for the straightforward integration of the $_pF_{p+1}$ functions arising from the 
logarithmic-trigonometric integrals involved.

{\bf Lemma 5}.  For $p \neq 0$, $q \neq -1$, and $(q+1)/p \neq -1$,
$$\int \kappa^q ~_3F_4(a_1,a_2,a_3;b_1,b_2,b_3,c;-\kappa^p/4)d\kappa$$
$$={\kappa^{q+1} \over {q+1}} ~_4F_5\left(a_1,a_2,a_3,{{q+1} \over p};b_1,b_2,b_3,c,
{{q+1} \over p}+1;- {\kappa^p \over 4}\right).  \eqno(3.29)$$

{\it Proof}. 
We have
$$\int \kappa^q ~_3F_4(a_1,a_2,a_3;b_1,b_2,b_3,c;-\kappa^p/4)d\kappa
=\sum_{\ell=0}^\infty {{(a_1)_\ell (a_2)_\ell(a_3)_\ell} \over {(b_1)_\ell (b_2)_\ell(b_3)_\ell}}{1 \over {(c)_\ell}}\left(-{1 \over 4}\right)^\ell {1 \over {\ell!}}\int \kappa^{p\ell+q} d\kappa$$
$$=\sum_{\ell=0}^\infty {{(a_1)_\ell (a_2)_\ell(a_3)_\ell} \over {(b_1)_\ell(b_2)_\ell(b_3)_\ell}}{1 \over {(c)_\ell}}\left(-{1 \over 4}\right)^\ell {1 \over {\ell!}}{{\kappa^{p\ell+q+1}} \over {(p\ell+q+1)}}.  \eqno(3.30)$$
We then use
$${{q+1} \over {p\ell+q+1}}={{\left({{q+1} \over p}\right)_\ell} \over {\left({{q+1} \over p}+1\right)_\ell}}, \eqno(3.31)$$ 
to complete the Lemma.

We then call out a special case at $c=(q+1)/p$. 
\newline{\bf Corollary 2}.  For $p \neq 0$,
$$\int \kappa^q ~_3F_4\left(a,a,a;b,b,b,{{q+1} \over p};-{{\kappa^p} \over 4}\right)d\kappa$$
$$={\kappa^{q+1} \over {q+1}} ~_3F_4\left(a,a,a;b,b,b,
{{q+1} \over p}+1;- {\kappa^p \over 4}\right).  \eqno(3.32)$$ 
In particular, this result applies with $p=q=2$ when integrating the
function of (3.6) with respect to $a$.

{\bf Lemma 6}.  (Leading asymptotic form of special $_pF_{p+1}$ functions)
For $z \to \infty$ we have (a)
$$_2F_3\left(1,1;2,2,{3 \over 2};-z\right) \sim {1 \over z}\left(\gamma+\ln 2+{1 \over 2}\ln z\right), \eqno(3.33)$$
(b)
$$_3F_4\left(1,1,1;2,2,2,{5 \over 2};-z\right) \sim {1 \over z}\left[6(1-\gamma)+3\gamma^2-{\pi^2 \over 4}-6\ln 2+6\gamma \ln 2+3\ln^2 2\right]$$
$$+{3 \over z}(\gamma-1+\ln 2)\ln z+{3 \over 4}{{\ln^2 z} \over z},  \eqno(3.34)$$
and (c)
$$_4F_5\left(1,1,1,1;2,2,2,2,{5 \over 2};-z\right) \sim {1 \over z}\left[12(\gamma-1)
-6\gamma^2+2\gamma^3+{\pi^2 \over 2}-{\gamma \over 2}\pi^2-6\ln^2 2+6\gamma \ln^2 \right. $$
$$+2\ln^3 2+6 \ln 2(\gamma-1)\ln z+3 \ln^2 2 \ln z-{3 \over 2} \ln^2 z+{3 \over 2}(\gamma+ \ln 2) \ln^2 z+\ln^3 z$$
$$\left.+(6(1-\gamma)+3\gamma^2-{\pi^2 \over 4})\ln(4z)+4\zeta(3)\right].  \eqno(3.35)$$

{\it Proof}.  (a) This follows immediately from the definition of Ci in (3.1) and
the relation (3.2).  However, we use a more general procedure based upon the Barnes
integral representation of $_pF_q$ (\cite{paris}, Section 2.3).  We have
$$_2F_3\left(1,1;2,2,{3 \over 2};-z\right)={1 \over {2\pi i}}\int_{-i\infty}^{i\infty}
\Gamma(-s)\Gamma(s+1)g(s)z^s ds, \eqno(3.36)$$
where the path of integration is a Barnes contour, indented to the left of the origin
but staying to the right of $-1$, and
$$\Gamma(s+1)g(s)=\Gamma\left({3 \over 2}\right){{\Gamma^2(s+1)} \over {\Gamma^2(s+2)}}
{1 \over {\Gamma(s+3/2)}}=\Gamma\left({3 \over 2}\right){1 \over {(s+1)^2\Gamma(s+3/2)}}
.  \eqno(3.37)$$
The contour can be thought of as closed in the right half plane, over a semicircle
of infinite radius.  We then move the contour to the left of $s=-1$, picking up the
residue there.  
We have
$${d \over {ds}}{{\Gamma(-s)} \over {\Gamma(s+3/2)}}=-{{\Gamma(-s)} \over {\Gamma(s+3/2)}}[\psi(-s)+\psi(s+3/2)], \eqno(3.38)$$ 
and $\psi(1/2)=-\gamma-2\ln 2$.  Therefore, we have for $s$ near $-1$
$$\Gamma\left({3 \over 2}\right){{\Gamma(-s)} \over {\Gamma(s+3/2)}}={1 \over 
2}+(\gamma+\ln 2)(s+1)+O[(s+1)^2], \eqno(3.39)$$
with $z^s=1/z+\ln z(s+1)/z+O[(s+1)^2]$ and find
$$\Gamma\left({3 \over 2}\right) \mbox{Res}_{s=-1} ~{{\Gamma(-s)z^s} \over {(s+1)^2 \Gamma(s+3/2)}}={1 \over z}\left(\gamma+\ln 2+{1 \over 2}\ln z\right), \eqno(3.40)$$
giving (3.25).

(b) We proceed similarly, with
$$ _3F_4\left(1,1,1;2,2,2,{5 \over 2};-z\right)={1 \over {2\pi i}} \int_{-i\infty}^{i\infty} \Gamma(-s)\Gamma(s+1)g(s)z^s ds, \eqno(3.41)$$
and
$$\Gamma(s+1)g(s)=\Gamma\left({5 \over 2}\right){{\Gamma^3(s+1)} \over {\Gamma^3(s+2)}}{1 \over {\Gamma(s+5/2)}}.  \eqno(3.42)$$
Again using an expansion like (3.39) and evaluating the residue at $s=-1$ gives (3.34).

(c)  Similarly, we use
$$ _4F_5\left(1,1,1,1;2,2,2,2,{5 \over 2};-z\right)={1 \over {2\pi i}} \int_{-i\infty}^{i\infty} \Gamma(-s)\Gamma(s+1)g(s)z^s ds, \eqno(3.43)$$
with
$$\Gamma(s+1)g(s)=\Gamma\left({5 \over 2}\right){{\Gamma^4(s+1)} \over {\Gamma^4(s+2)}}
{1 \over {\Gamma(s+5/2)}}=\Gamma\left({5 \over 2}\right){1 \over {(s+1)^4\Gamma(s+5/2)}},
\eqno(3.44)$$
and compute 
$$\Gamma\left({5 \over 2}\right) \mbox{Res}_{s=-1} ~{{\Gamma(-s)z^s} \over {(s+1)^4 \Gamma(s+3/2)}}.$$

{\it Remarks}.  We note that results like part (b) of this Lemma allows us to find the limit as $y \to \infty$ for the integral of Lemma 2.

We have presented the algebraic part of the asymptotic form of the $_pF_{p+1}$ functions, that is the leading portion.  The higher order terms
come from the exponential expansion of these functions, and they are infinite
in number.  According to (\cite{paris}, Section 2.3, p. 57) this expansion
has the form
$$E(-z)=(2iz^{1/2})^\theta e^{2i\sqrt{z}}\sum_{k=0}^\infty {A_k \over
{(2iz^{1/2})^k}}, \eqno(3.45)$$
where $\theta=1-p-\alpha$ and $A_0=(2\pi)^{-1/2}2^{-1/2-\theta}$.  Here,
$\alpha$ is either $3/2$ (for the $p=2$ case) or otherwise $5/2$.  The 
other coefficients $A_k$ come from an inverse factorial expansion of the 
function $g(s)$,
$$g(s)=2\cdot 4^s\left[\sum_{j=0}^\infty {A_j \over {\Gamma(2s+1-\theta+j)}}
+{{O(1)} \over {\Gamma(2s+1-\theta+M)}}\right].  \eqno(3.46)$$  
The full asymptotic expansion of the $_pF_{p+1}$ functions includes the
algebraic portion $H(-z)$ given in the Lemma, together with (3.42), as
$$ ~_pF_{p+1}\left(1,1,\ldots,1;2,2,\ldots,2,{5 \over 2};-z \right) \sim
E(-z)+E(z)+H(-z).  \eqno(3.47)$$


{\bf Lemma 7}. (Alternative integral representation)
$$_3F_4\left(1,1,1;2,2,2,{5 \over 2};-{\kappa^2 \over 4}\right)={3 \over 2}{1 \over \kappa^3}\int_0^\infty x^2[-\kappa \cos(e^{-x/2}\kappa)+e^{x/2}\sin(e^{-x/2}\kappa)]dx.
\eqno(3.48)$$

{\it Proof}.  This follows by using the special case Laplace transform
$$\int_0^\infty x^2 e^{-rx}dx={{\Gamma(3)} \over r^3}.  \eqno(3.49)$$
Then, with the integral being absolutely convergent, we may interchange summation and integration in
$$_3F_4\left(1,1,1;2,2,2,{5 \over 2};-{\kappa^2 \over 4}\right)={1 \over 2} \sum_{j=0}^
\infty \int_0^\infty x^2 e^{-(j+1)x} {1 \over {(5/2)_j}}\left(-{\kappa^2 \over 4}\right)^j {1 \over {j!}}dx, \eqno(3.50)$$
and the result follows.  

We mention the connection between integrals expressible as certain differences of the
confluent hypergeometric function $_1F_1$, differences of the incomplete gamma function
$\gamma(x,y)$, and a $_1F_2$ function.  We have
\newline{\bf Lemma 8}.  We have for Re $\mu>-1$
$$\int_0^1 x^{\mu-1} \sin(ax)dx=-{i \over {2\mu}}[~_1F_1(\mu;\mu+1;ia)- ~_1F_1(\mu;\mu+1;-ia)]$$
$$=-{i \over 2}(ia)^{-\mu}[\gamma(\mu,-ia)-(-1)^\mu \gamma(\mu,ia)]$$
$$={a \over {\mu+1}} ~_1F_2\left({{1+\mu} \over 2};{3 \over 2},{{3+\mu} \over 2};-{a^2\over 4}\right).  \eqno(3.51)$$

We note that by logarithmic differentiation with respect to $\mu$ this result 
generates a family of logarithmic sine integrals.  I.e., for Re $\mu>-1$,
$$\int_0^1 x^{\mu-1} \sin(ax) \ln^k x ~dx=\left({\partial \over {\partial \mu}}\right)^k \int_0^1 x^{\mu-1} \sin(ax)dx.  \eqno(3.52)$$

{\it Proof}.  The first line of (3.51) is from (\cite{grad}, p. 420, 3.761.1).
The second line follows from the relation (\cite{grad}, p. 1063)
$$\gamma(\alpha,x)={1 \over \alpha}x^\alpha ~_1F_1(\alpha;\alpha+1;-x).  \eqno(3.53)$$
For the third line we write
$$~_1F_1(\mu;\mu+1;ia)-~_1F_1(\mu;\mu+1;-ia)=\sum_{j=0}^\infty {{(\mu)_j} \over {(\mu+1)
_j}}{{(ia)^j} \over {j!}} [1-(-1)^j]$$
$$=2i \sum_{m=0}^\infty {{(\mu)_{2m+1}} \over {(\mu+1)_{2m+1}}}{a^{2m+1} \over {(2m+1)!}}
. \eqno(3.54)$$
The Lemma is now completed with the use of the duplication formula, as at the beginning 
of the proof of Lemma 2, and the relation
$${{(\mu)_{2m+1}} \over {(\mu+1)_{2m+1}}}={\mu \over 2}{1 \over {(\mu/2+m+1/2)}}
={\mu \over {2(\mu+1)}}{{\left({{1+\mu} \over 2}\right)_m} \over {\left({{3+\mu} \over 2}
\right)_m}}.  \eqno(3.55)$$

\medskip
\centerline{\bf Discussion}
\medskip

We mention how hypergeometric summatory representations may be obtained
for the constants $\gamma_k(a)$.  When $k=0$ we must recover from (2.2) the known representations (e.g., \cite{edwards}, p. 107)
$$\ln \Gamma(a)=\left(a-{1 \over 2}\right)\ln a -a+{1 \over 2}\ln (2\pi)-\int_0^\infty
{{P_1(t)} \over {t+a}}dt, \eqno(4.1)$$
and
$$\psi(a)=-\gamma_0(a)=\ln a-{1 \over {2a}}+\int_0^\infty {{P_1(t)} \over {(t+a)^2}}dt, \eqno(4.2)$$
leading to, for Re $a>0$,
$$\psi(a)=-\gamma_0(a)=\ln a-{1 \over {2a}}+\sum_{j=1}^\infty \left[2\cos(2\pi ja)\mbox{Ci}(2\pi ja)-\sin(2\pi ja)[\pi-2\mbox{Si}(2\pi ja)]\right].  \eqno(4.3)$$

Matters are more interesting, and challenging, for $k \geq 1$.

From (2.2) we have
$$\zeta'(s,a)=-{{a^{-s}\ln a} \over 2}-{{a^{1-s}\ln a} \over {s-1}}
-{a^{1-s} \over {(s-1)^2}}$$
$$+\int_0^\infty {{P_1(x)} \over {(x+a)^{s+1}}}[s\ln(x+a)-1]dx, \eqno(4.4)$$
so that
$$-\gamma_1(a)=-{{\ln a} \over {2a}}+{{\ln^2 a} \over 2}+\int_0^\infty {{P_1(x)} \over {(x+a)^2}}[\ln(x+a)-1]dx. \eqno(4.5)$$
Again the Fourier series (2.1), applies so that the term
$$\int_0^\infty {{P_1(x} \over {(x+a)^2}}[\ln(x+a)-1]dx
=-{1 \over \pi}\sum_{n=1}^\infty {1 \over n}\int_0^\infty {{\sin(2\pi n x)} \over
{(x+a)^2}}[1-\ln (x+a)]dx$$
$$=-{1 \over \pi}\sum_{n=1}^\infty {1 \over n}\int_a^\infty {{\sin[2\pi n (x-a)]} \over
x^2}(1-\ln x)dx$$
$$=-{1 \over \pi}\sum_{n=1}^\infty {1 \over n}\int_a^\infty{{[\sin(2\pi n x)\cos 2\pi na -\cos(2\pi nx)\sin 2\pi n a]} \over x^2} (1-\ln x)dx. \eqno(4.6)$$
A special case occurs at $a=1/2$, when $\zeta(s,1/2)=(2^s-1)\zeta(s)$,
implying that $\gamma_1(1/2)=\gamma_1-2\gamma \ln 2-\ln^2 2$.
We have from (4.4) and (4.5)
$$-\gamma_1(1/2)=\ln 2+{1 \over 2}\ln^2 2-{1 \over \pi}\sum_{n=1}^\infty 
{{(-1)^n} \over n} \int_{1/2}^\infty {{\sin(2\pi n x)} \over x^2}(1-\ln x)
dx.  \eqno(4.7)$$
Here
$$\int_{1/2}^\infty {{\sin(2\pi n x)} \over x^2}dx=-\kappa \mbox{Ci} (\kappa/2)+2\sin(\kappa/2), \eqno(4.8)$$
with the last term vanishing when $\kappa=2\pi n$.  The logarithmic
integral in (4.6) can be decomposed as $\int_{1/2}^\infty =\int_{1/2}^1+
\int_1^\infty$ and Lemma 2 and (2.7) can be applied respectively.  This
leads to a hypergeometric summatory expression for $\gamma_1(1/2)$.
 
Lastly, we mention that our methods for treating logarithmic integrals may be
applied to other functions, including the Bessel functions of the first kind $J_n$.
These integrals may be of the form
$$\int_1^\infty {{J_n(ax)} \over x^j}\ln^k(bx)dx,$$
and in particular for $n=1$ we expect analogies with our previous trigonometric logarithmic integrals.  However, we do not pursue that topic here.

\bigskip
\centerline{\bf Summary}
\medskip

We have presented summations over particular $_pF_{p+1}$ generalized
hypergeometric functions and mathematical constants giving the Stieltjes
constants $\gamma_k$.  
We have described the asymptotic form of the $_pF_{p+1}$ functions with
repeated parameters, and indeed the extended asymptotics may be used to
accelerate the convergence of the hypergeometric-based summations.
We have considered certain logarithmic-sine integrals.  We showed that
these are essentially certain iterated integrals with the cosine integral
Ci at their base. Our results generalize significantly simpler 
summatory expressions for the Euler constant $\gamma$ in terms of
evaluations of Ci or the sine integral Si.



\pagebreak
\centerline{\bf Appendix:  Logarithmic integrals}
\medskip

The following two integrals are given in the Table \cite{grad} (p. 647):
$$\int_0^\infty \mbox{Ci}(x)\ln x ~dx={\pi \over 2}, \eqno(A.1)$$
and
$$\int_0^\infty \mbox{si}(x)\ln x ~dx=\gamma+1. \eqno(A.2)$$
Here,
$$\mbox{si}(x) \equiv -\int_x^\infty {{\sin t} \over t}dt.  \eqno(A.3)$$

We show that these two integrals follow as special cases of the following.
{\newline \bf Proposition A1}.  For $a>0$,
$$\int_0^\infty e^{-at}\cos t(1-\ln t)dt={1 \over {2(1+a^2)}}\left\{2 \cot^{-1}a+a \left[\ln\left(1+{1 \over a^2}\right)+2(1+\gamma+\ln a)
\right]\right\}, \eqno(A.4)$$
and
$$\int_0^\infty e^{-at}\sin t(1-\ln t)dt={1 \over {2(1+a^2)}}\left[-2a \cot^{-1}a+\ln\left(1+{1 \over a^2}\right)+2(1+\gamma+\ln a)\right].  
\eqno(A.5)$$

We first observe the special cases under formal interchange of integrations
$$\int_0^\infty \mbox{Ci}(x)\ln x ~dx=-\int_0^\infty \ln x \int_x^\infty
{{\cos t} \over t}dt dx$$
$$``="-\int_0^\infty {{\cos t} \over t}\int_0^t \ln x ~dx dt$$
$$=\int_0^\infty \cos t(1-\ln t)dt, \eqno(A.6)$$
and similarly
$$\int_0^\infty \mbox{si}(x)\ln x ~dx=-\int_0^\infty \ln x \int_x^\infty
{{\sin t} \over t}dt dx$$
$$``="-\int_0^\infty {{\sin t} \over t}\int_0^t \ln x ~dx dt$$
$$=\int_0^\infty \sin t(1-\ln t)dt. \eqno(A.7)$$
Indeed the integrals (A.6) and (A.7) are not convergent.  However,
the $a \to 0$ limit in the Proposition gives the integrals (A.1) and (A.2).

{\it Proof}.  We just prove (A.4), as (A.5) goes similarly.  From the
Gamma function integral
$$\int_0^\infty e^{-at} t^{z-1}dt=a^{-z} \Gamma(z) \eqno(A.8)$$
we have
$$\int_0^\infty e^{-at} t^{z-1}\ln t ~dt={\partial \over {\partial z}}a^{-z} \Gamma(z)$$
$$=a^{-z}\Gamma(z)[\psi(z)-\ln a]. \eqno(A.9)$$
We recall the relation $\psi(n+1)=H_n-\gamma$, where $H_n$ is the $n$th
harmonic number, and then for integers $j \geq 0$
$$\int_0^\infty e^{-at} t^{2j} \ln t ~dt=a^{-2j+1}\Gamma(2j+1)[\psi(2j+1)
-\ln a]$$
$$=a^{-(2j+1)} (2j)![H_{2j}-\gamma-\ln a].  \eqno(A.10)$$
Then integrating term-by-term we have
$$\int_0^\infty e^{-at}\cos t\ln t~dt=\sum_{j=0}^\infty {{(-1)^j} \over {(2j)!}}\int_0^\infty e^{-at} t^{2j} \ln t ~dt$$
$$=\sum_{j=0}^\infty {{(-1)^j} \over {(2j)!}} a^{-(2j+1)} (2j)![H_{2j}-\gamma-\ln a]$$
$$=-{{(\gamma+\ln a)} \over {1+a^2}}+\sum_{j=0}^\infty (-1)^j H_{2j} 
a^{-(2j+1)}.  \eqno(A.11)$$
The latter sum may be performed with the substitution of an integral
representation for the digamma function, or by manipulating the generating
function 
$$\sum_{n=0}^\infty (-1)^n H_n z^n =-{{\ln(1+z)} \over {1+z}}, \eqno(A.12)$$
and recalling that $\cot^{-1}z=(1/(2i))\ln[(1+iz)/(1-iz)]$.
We have
$$\sum_{n=0}^\infty [(-1)^n+1]H_nz^n=2\sum_{m=0}^\infty H_{2m}z^{2m}=
-{{\ln(1+z)} \over {1+z}}+{{\ln(1-z)} \over {z-1}}.  \eqno(A.13)$$
We then put $z=ia$ to obtain the sum of (A.11).  Using the elementary
integral $\int_0^\infty e^{-at} \cos t dt=a/(1+a^2)$ completes the
Proposition.

\pagebreak


\begin{thebibliography}{99}
\bibitem{ftnote}
{It appears that in (5) and several later equations in Sections 2-4 in \cite{matsuoka2} a factor of $n!$ is missing.}
\end{thebibliography}
\end{document}